\documentclass[sigconf]{acmart}

\setcopyright{none}
\setcopyright{rightsretained}

\makeatletter
\renewcommand\@formatdoi[1]{\ignorespaces}
\makeatother

\acmISBN{}

\acmConference[KDD MLG 2018]{14th International Workshop on Mining and Learning with Graphs}{Aug 20, 2018}{London, United Kingdom}
\acmYear{2018}
\copyrightyear{2018}
\acmPrice{}

\renewcommand\footnotetextcopyrightpermission[1]{} 
\usepackage[utf8]{inputenc}
\usepackage{color}

\usepackage{listings}
\usepackage{graphicx}
\usepackage{subcaption}
\usepackage{amsmath}
\usepackage{booktabs} 
\usepackage{hyperref}

\begin{document}

\title[From clusters to queries]{From clusters to queries: exploiting uncertainty in the modularity landscape of complex networks}

\author{James P Gilbert}
\affiliation{%
  \institution{Synthetic Biology Research Centre}
  \streetaddress{University of Nottingham}
  \city{Nottingham}
  \country{United Kingdom}
  \postcode{NG7 2RD}
}
\email{james.gilbert@nottingham.ac.uk}

\author{Jamie Twycross}
\affiliation{%
  \institution{School of Computer Science}
  \streetaddress{University of Nottingham}
  \city{Nottingham}
  \country{United Kingdom}
  \postcode{NG8 1BB}
}
\email{jamie.twycross@nottingham.ac.uk}

\begin{abstract}
Uncovering latent community structure in complex networks is a field that has received an enormous amount of attention.
Unfortunately, whilst potentially very powerful, unsupervised methods for uncovering labels based on topology alone has been shown to suffer from several difficulties.
For example, the search space for many module extraction approaches, such as the modularity maximisation algorithm, appears to be extremely glassy, with many high valued solutions that lack any real similarity to one another.
However, in this paper we argue that this is not a flaw with the modularity maximisation algorithm but, rather, information that can be used to aid the context specific classification of functional relationships between vertices.
Formally, we present an approach for generating a high value modularity consensus space for a network, based on the ensemble space of locally optimal modular partitions.
We then use this approach to uncover latent relationships, given small query sets.
The methods developed in this paper are applied to biological and social datasets with ground-truth label data, using a small number of examples used as seed sets to uncover relationships.
When tested on both real and synthetic datasets our method is shown to achieve high levels of classification accuracy in a context specific manner, with results comparable to random walk with restart methods.
\end{abstract}

\maketitle

\section{Introduction}
A fundamental issue at the heart of machine learning methods applied to large scale datasets is the ability to correctly identify classes of related objects in an unsupervised manner.
In network science, this methodology is often referred to as \textit{community} detection \cite{fortunato2016community}.
Many algorithms exist to solve this problem \cite{fortunato2016community}, yet initial starting conditions or different optimisation strategies may result in conflicting results even when the same objective function is being maximised \cite{good2010performance}.
In this paper, we develop an intuitive method to use the uncertainty amongst a high number of near optimal solutions to measure context sensitive relationships between small sets of labelled vertices.
This approach can be based on labels that are not first order neighbours to find other potentially related vertices.

The use of community detection is widespread.
For example, a core goal in systems biology is to characterise the function and functional relationships between genes, proteins or metabolites within a larger network \cite{girvan2002community}.
In many situations, only the role of a small number of genes is known, with much of the annotation for a given organism being computed through naive homology information that ignores the role of a gene within a wider context.
The advent of high throughput experimental datasets has allowed the construction of proteome scale networks, leading to the observation of non-trivial topological properties such as densely connected clusters \cite{ArabidopsisConsortium2011}.
These densely connected clusters are widely believed to be associated with specific function, such as multi-protein complexes or biochemical pathways.

As a form of unsupervised machine learning, module extraction methods largely focus on optimising some objective function with the goal of finding meaningful clusterings.
Perhaps the most popular of these methods is that of modularity maximisation \cite{newman2004}, which seeks to find the most unexpected partition of a graph with respect to a given null model.
Overlapping methods have recently been applied to this problem in both \textit{crisp} \cite{ahn2010link, lancichinetti2011finding} and \textit{fuzzy} \cite{gregory2011fuzzy} based algorithms which have been widely used to uncover latent relationships without labelling schemes.
In previous work, we found that most of these methods have significant disagreement when evaluated in a practical context \cite{gilbert2015probabilistic}.

The number and size of communities is, generally, not known \textit{a priori}, and the problem has been shown to be NP-hard \cite{npHardModularity}.
The work of Good \textit{et al.} \cite{good2010performance} recently highlighted that the popular modularity maximisation algorithm has a highly ``glassy'' search space.
In essence, for real, heterogeneous networks there are many locally optimal partitions that bear little resemblance to each other by measure of mutual information.
This allows greedy optimisation algorithms \cite{blondel2008fast} to trivially find solutions that score extremely high values of modularity.
In order to solve this issue, certain approaches use a consensus based approach to clustering, combining many high value partitions into a given median partition \cite{lancichinetti2012consensus}.

However, in this paper, we do not seek to find a single ``best'' partition, whether overlapping or not and our objective is not to uncover labels but to use limited and small sets of labels to give the notion of a membership score to some grouping.
Instead, we use the large number of highly modular solutions to form the index for a search query system.
In essence, this is a method of semi-supervised learning that attempts to find items related given labelled sets of vertices using topology alone.
Each high value partition can be treated as information about the relationships between vertices.
That is to say, there is not a single, correct view of the underlying community structure to a network, but rather, many different context dependent definitions.
As the objective of community detection approaches is to relate information, it is assumed that some labelled meta-data can be used to find unlabelled, potentially related vertices.

More formally, the problem tackled in this paper can be formulated as follows:
Given a graph made up of vertices and edges $G = (V, E)$, and a query set $S \subset V$, we ask the question; \textit{How well is a given vertex, $i \notin S$ related to $S$?}

We propose an algorithm that pre-computes an index of clusterings for a given complex network, based on the fast greedy Louvain algorithm \cite{blondel2008fast}, used in a distributed manner.
The detected clusters then form the basis of a search algorithm that allows one to compute the relatedness of nodes to a given query set.
The querying method is a polynomial time algorithm that could be trivially adapted to form the basis of many user facing applications.
This approach is then applied to synthetic benchmark networks with known, ground-truth labels as well as social and protein-protein interaction networks with high quality ground truth label sets.

\begin{figure}[t]
    \includegraphics[width=0.48\textwidth]{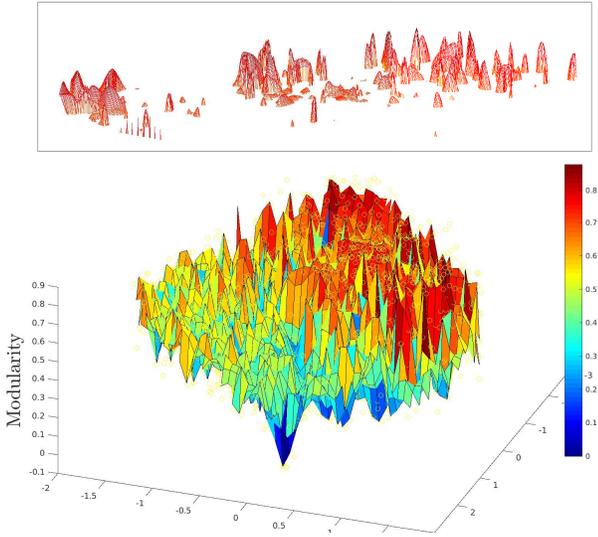}
    \caption{Modularity search space of an \textit{E. coli} metabolic network.
 Distance between partitions is calculated using variation of information \cite{meilua2003comparing} and dimensionality reduction is performed using curvilinear component analysis \cite{demartines1997curvilinear}.
  The inset (top) demonstrates the landscape of the high modularity partitions.
  Figure generated with the software of Good \textit{et al.} \cite{good2010performance}.}
    \label{fig:modular_search_space}
\end{figure}

\section{Exploiting the modularity query space}
To characterise latent community structure, one of the most popular approaches is to use modularity maximisation given by the equation \cite{newman2004}
\begin{equation}\label{eq:modularity}
  Q = \frac{1}{2m}\sum_{i,j} \left[A_{ij} - \frac{k_i k_j}{2m}\right]\delta(c_i, c_j),
\end{equation}
where $m$ is the number of edges in the network, $A_{ij}$ is the binary variable indicating the adjacency of nodes $i$ and $j$, $k_i$ is the degree of a vertex, $c_i$ indicates the community of a given vertex and $\delta(c_i, c_j)$ is the Kronecker delta such that $\delta(c_i, c_j) = 1$ if $c_i = c_j$ and $0$ otherwise.
As a combinatorial optimisation problem, there are many different algorithmic approaches to finding high values of $Q$.

The work by \cite{good2010performance} forms the basis of the motivation of the approach taken here.
In this study, the authors discovered that the modularity search space for many real-world networks contains a huge number of high value solutions.
Each of these solution partitions are extremely close to the global maxima, making it both difficult to find the optimal value of $Q$ and difficult to argue that the highest scoring partition is the ``true'' community structure.
This fact is visually demonstrated with the software from \cite{good2010performance} in Figure \ref{fig:modular_search_space}, which shows the modularity search space of an \textit{E. coli} metabolic network reconstruction \cite{GuimeraNature2005}.
The similarity between the partitions is compared with the variation of information measure \cite{meilua2003comparing} and dimensionality of the space is reduced with curvilinear component analysis \cite{demartines1997curvilinear} \footnote{The reader should note that axis on these plots are a result of the dimensionality reduction performed by curvilinear component analysis \cite{demartines1997curvilinear} and, therefore, have no natural interpretation.}.

\begin{figure}[t]
    \centering
    \includegraphics[width=0.45\textwidth]{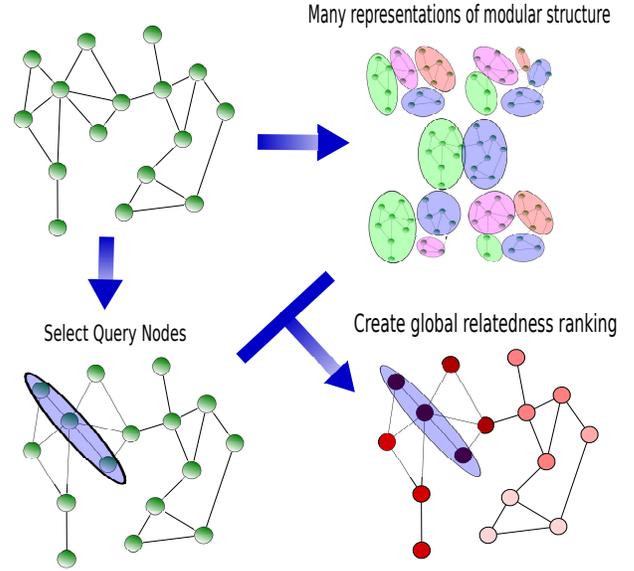}
    \caption{Outline of the proposed approach to querying networks by using multiple, high quality representations of modular networks.}
    \label{fig:algorithm_outline}
\end{figure}

In this work, we consider each high value partition to be information about latent relationships between vertices inferred through the topology of the network.
This approach, in and of itself is not unique, as there have been previous approaches that use the consensus of an ensemble of clusters to create high quality overlapping clustering of networks \cite{lancichinetti2012consensus}.
Such an approach, whilst well principled, is a context insensitive view of the modular structure of a graph.

The objective, then, is to use the disagreement between the set of highly modular partitions as information; that is to say, to infer the probability that sets of vertices are contained within the same cluster.
Whilst methods based on simulated annealing can be used to guarantee full coverage of the network, the following section describes a method adapted from the greedy agglomerative Louvain algorithm \cite{blondel2008fast}.

\subsection{Algorithm outline}
A broad outline of the proposed method is presented in Figure \ref{fig:algorithm_outline}.
In essence, the objective is to use multiple modular representations of a given dataset to generate a relatedness score for a given set of query vertices.

In order to cover the space of high modularity partitions, randomly generated starting partitions are computed with a random cut set.
To achieve this, each edge is either placed inside the cut set or not as the result of an independent Bernoulli trial. 
Each random partition is used as the starting partition for the greedy Louvain process.
In principle, any search exploration procedure, such as simulated annealing \cite{good2010performance} could be used.
The Louvain algorithm was selected as it is fast, running in $O(n \log n)$ time complexity \cite{blondel2008fast}, and because it is a greedy algorithm and is guaranteed to stop after finding a locally maximal solution.

The Louvain algorithm is conceptually very simple; starting from a random partition, clusters are agglomerated if the merge results in a positive change in modularity $\Delta Q$.
When no possible moves result in an increase in modularity the algorithm has found a local optima.

The index \textit{coverage} is directly proportional to the number of starting random partitions.
A full coverage index could be considered as every locally optimal partition.
Given that there is no free lunch and it is impossible to know every solution, one can only ensure a full coverage index through an exhaustive search over the $2^m$ possible starting cut sets, where m is the number of edges in the graph.
Consequently, the approach taken here is to use a large but not exhaustive subset of the possible solutions using a suitably large space of 2000 solutions for the networks studied in this paper
\footnote{Initial results indicate that a significantly smaller space of partitions may still yield high quality results but we note that further work is required to asses the best number of partitions in a practical context.}.

\subsection{Measuring the quality of relationships} \label{sec:expansion}
Given a query of vertices, the relatedness to other vertices in a network is quantifiable by the fraction of times they are clustered with the query set, given the set of high quality partitions.
Formally, this can be expressed in terms of the \textit{expansion score} of a given vertex,
\begin{equation} \label{eq:mu_score}
\mu_i(S) = \frac{1}{|\mathcal{P}| |S| } \sum_{P \in \mathcal{P}} \sum_{j \in S} \delta(c^{p}_i, c^{p}_j),
\end{equation}
where $S$ denotes a query set, $P$ is a given partition in the space of all high quality partitions $\mathcal{P}$, $c^{p}_{i}$ indicates the cluster vertex $i$ is contained in within partition $P$ and
$\delta(u, v)$ is the Dirac delta function that equals 1 if $c^{p}_i$ and  $c^{p}_j$ are the same cluster and $0$ otherwise.
As a simple example, for a pair of vertices $i$ and $S$ such that $S = \{j\}$ we would consider $\mu_i$ to be the number of times $i$ and $j$ appear in the same cluster, given an ensemble of network clusterings.
We define $\mu_i(S)$ for all vertices in the network, including those in $S$.
However, for the case where $i$ is in $S$ we, instead, consider the value $\mu_i(S - i)$ to remove bias.

\section{Results}
\subsection{Cross-validation method}
\label{sec:cross_validation}
In this study we test a small number of labels that we intend to use in order to evaluate how well our method correctly generalises to discover unlabelled vertices.
We would like to test a significantly smaller number of seed nodes than two class classification methods used in previous studies, which use leave-one-out cross validation \cite{kohler2008walking}.
The cross validation procedure we devise is described as follows and depends on the size of the community and the number of initial seed labels being used.

For this work we would like to capture binary classification  performance, \textit{true positives (tp)}, \textit{true negatives (tn), false positives (fp)} and \textit{false negatives (fn)}, on our datasets of community memberships.
In order to generate the different sets for the cross validation we take each label set and generate unique sample sets of vertices from the known true positive labels.

As the seed label sets can be as small as 3 vertices, exhaustive cross validation was not possible for all labelling schemes.
Consequently, cross validation is either conducted on an exhaustive set of all possible $\binom{|S|}{s}$ unique labellings or 120 seed queries sampled randomly without replacement from the possible subsets \footnote{This is equivalent to $\binom{10}{3}$ combinations, given time constraints, an exhaustive sample would not be possible for the larger communities in this study.}, where $S$ is the set of gold standard true labels and $s$ is the size of the randomly selected seed sets.

As the selected seed sets can be contained within multiple communities, we consider the set of true positives not to be the community for which the seed set is randomly selected, but all communities for which that seed set is a subset of.
This is because the purpose of the approach presented within this paper is to distinguish between different communities in a context specific manner, if the overlap between two communities is represented by the seed nodes this should be considered in the tests.

It should therefore be noted that presented receiver operator characteristic (ROC) scores are dependent on community sizes.
We also note that we do not consider separate training and test sets in this study as the method does not use any examples when building the index space from partitions of the graph.

Formally, the steps for this procedure with a given label set $S$ are outlined as follows:
\begin{itemize}
 \item Generate up to 120 unique subsets of size $s$, randomly sampled without replacement (set $T$).
 \item For each test subset $\{\forall S_t \in T | S_t \subset S \wedge |S_t| = s \}$, generate the $\mu_i(S_t)$ score for all vertices in the network
 \item Exclude vertices in $S_t$ from the test
 \item Consider the true community membership of $S_t$ to be the true positive set.
 \item Consider each of the nodes in $V$ not in the community membership of $S_t$ to be the true negative set.
 \item Generate a network wide average ROC curve interpolated from all test subsets from all communities.
 \end{itemize}

In the case for the synthetic networks tested below, true labels are drawn from the known communities, with each tested in isolation.
For the real world networks tested the labels are considered in the same manner though many nodes have no assigned labels.
Every community is considered with different seed set sizes.
 
\subsection{Random walk with restart}
\label{sec:rwr}
In the following sections compare our method with the commonly used random walk with restart method as described in \cite{kohler2008walking}.
This simulates an infinite random walk with a fixed probability, $\alpha$ that a walker would teleport back to the initial seed nodes. 
Formally, the random walk with restart algorithm uses the form
\begin{equation} \label{eq:rwr}
 \vec{p}_{t+1} = (1 - \alpha) W \vec{p}_t + \alpha \vec{p}_0,
\end{equation}
where $W$ is the row normalised adjacency matrix of a graph, $\vec{p}_0$ is the initial walk vector such that $\vec{p}_0 = \frac{1}{|S|}$ for vertices in the seed set.
The random walk algorithm repeats equation \ref{eq:rwr} from $\vec{p}_0$ until the L1 norm between  $\vec{p}_t$ and  $\vec{p}_{t+1}$ converges to 0, simulating the steady state vector $\vec{p}_{\infty}$.
For the tests performed in this work we use a restart probability $\alpha = 0.25$.

\subsection{Synthetic networks}
In this section we test the method on benchmark networks constructed with a ground-truth community structure.
To evaluate how our method performs we use the undirected, unweighted LFR benchmark \cite{lancichinetti2008benchmark} in overlapping and non-overlapping forms.
We test the area under the ROC curve (AUC) scores for networks varying the mixing parameter (fraction of edges between communities) and the fraction of overlapping nodes.
In these tests, the community distribution is defined with a power law coefficient of $-1.0$, the degree distribution is defined with a power law coefficient of $-2.0$.
The number of nodes is stated in the text.

\subsubsection{Non-overlapping modules.}
\begin{figure*}[t]
    \centering
    \begin{subfigure}[b]{0.45\textwidth}
        \centering
        \includegraphics[width=\textwidth]{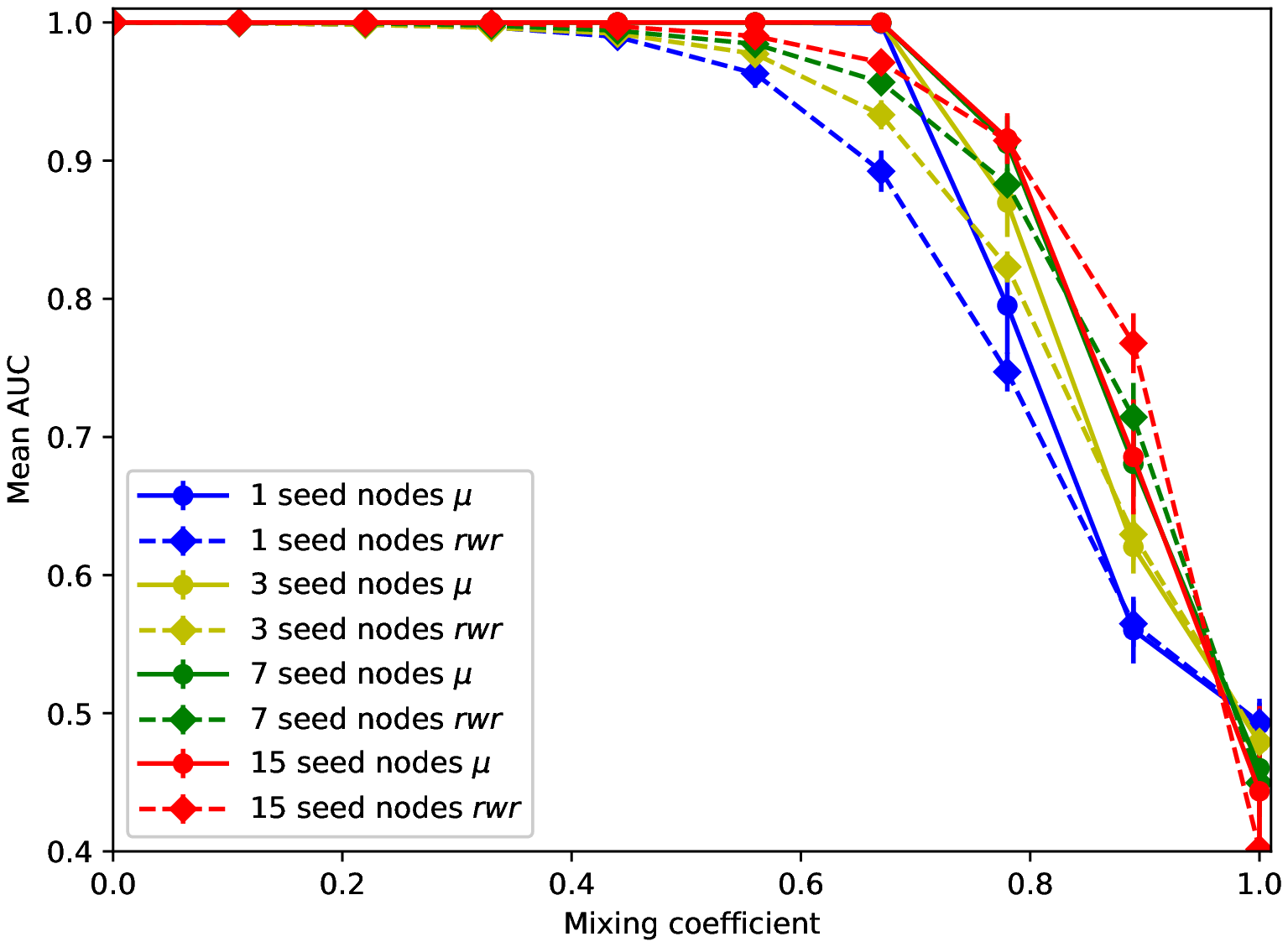}
        \caption{1000 nodes}
    \end{subfigure}
    \begin{subfigure}[b]{0.45\textwidth}
        \centering
        \includegraphics[width=\textwidth]{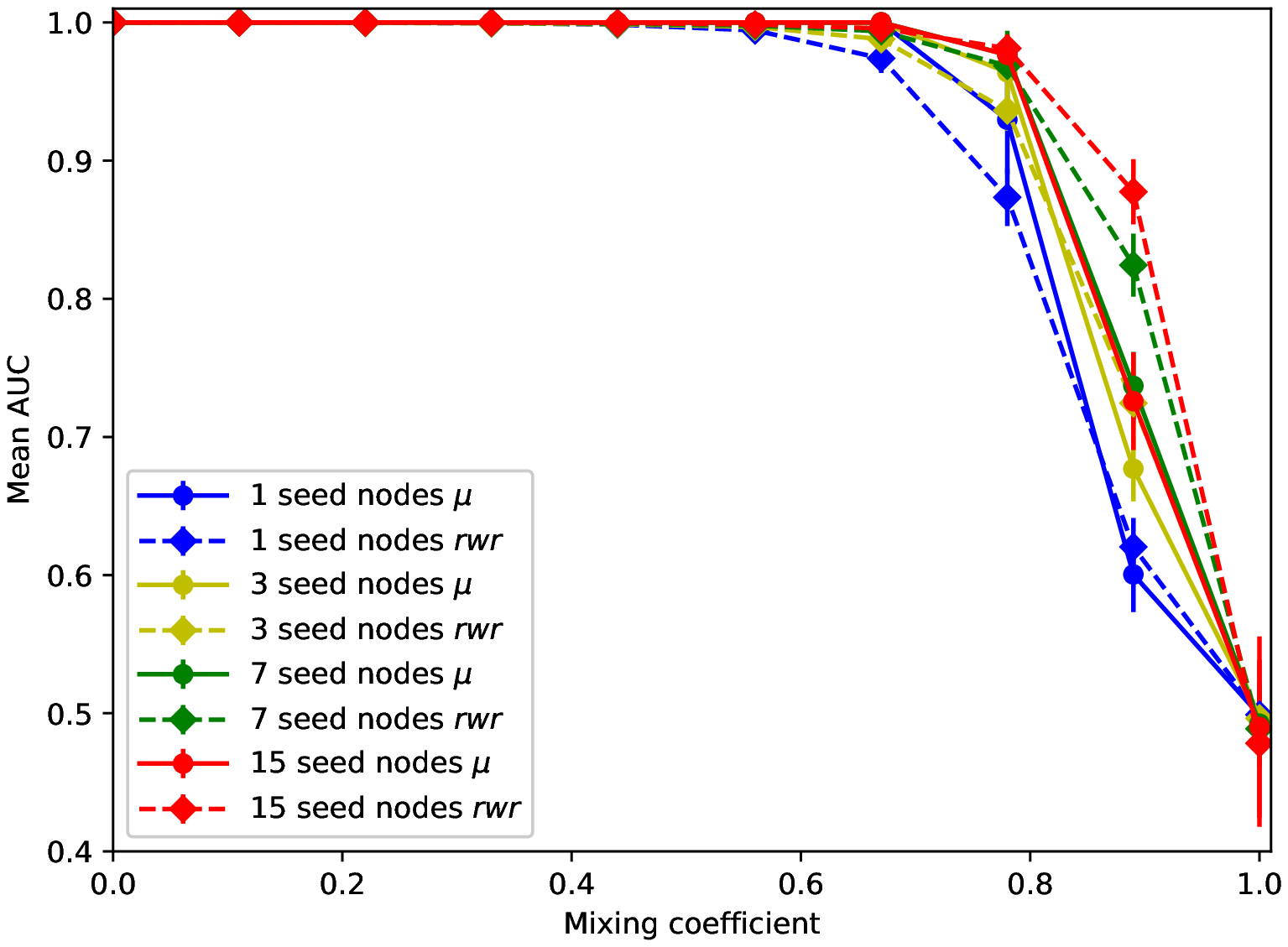}
        \caption{5000 nodes}
    \end{subfigure}
    \caption{Non-overlapping LFR community networks with varying seed nodes with 1000 and 5000 nodes.
     Data points represent mean AUC scores for all communities on 10 sampled networks at varying mixing coefficients.
     Error bars represent one standard deviation.}
     \label{fig:auc_no_overlap}
\end{figure*}

Figure \ref{fig:auc_no_overlap} represents tests on 10 networks with 1000 and 5000 nodes varying the mixing coefficient (number of edges between communities).
As one would expect, the prediction of the method drops off steeply where communities are less defined above a mixing coefficient of 0.6.
Using larger seed sets also improves accuracy with some prediction of true communities being possible at extremely high levels of mixing.
Overall, results are comparable to the random walk with restart method, with slightly improved performance with a smaller number of seed nodes.

\subsubsection{Overlapping modules.} 
\begin{figure*}[t]
    \centering
    \begin{subfigure}[b]{0.45\textwidth}
        \centering
        \includegraphics[width=\textwidth]{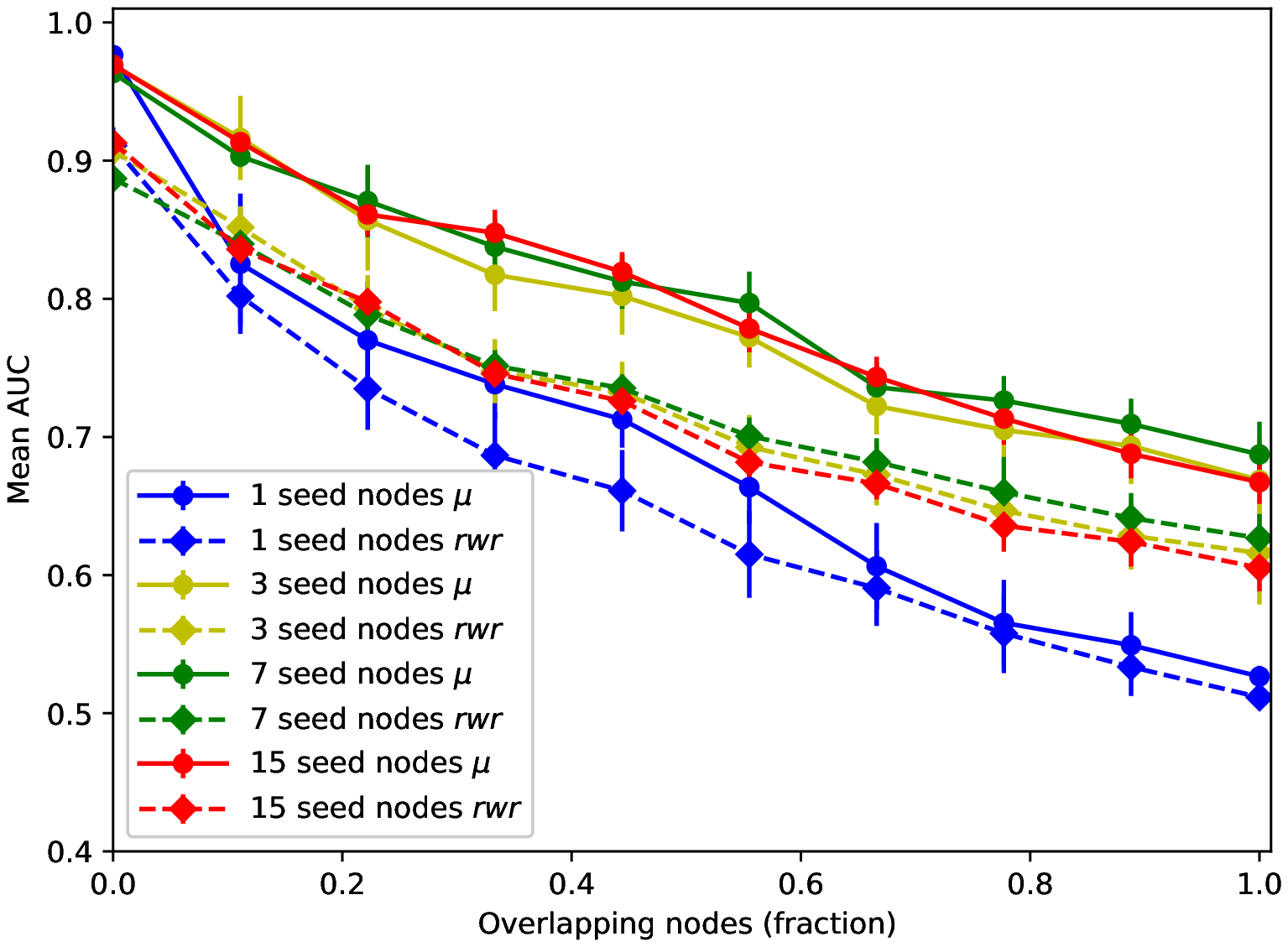}
        \caption{1000 nodes}
    \end{subfigure}
    \begin{subfigure}[b]{0.45\textwidth}
        \centering
        \includegraphics[width=\textwidth]{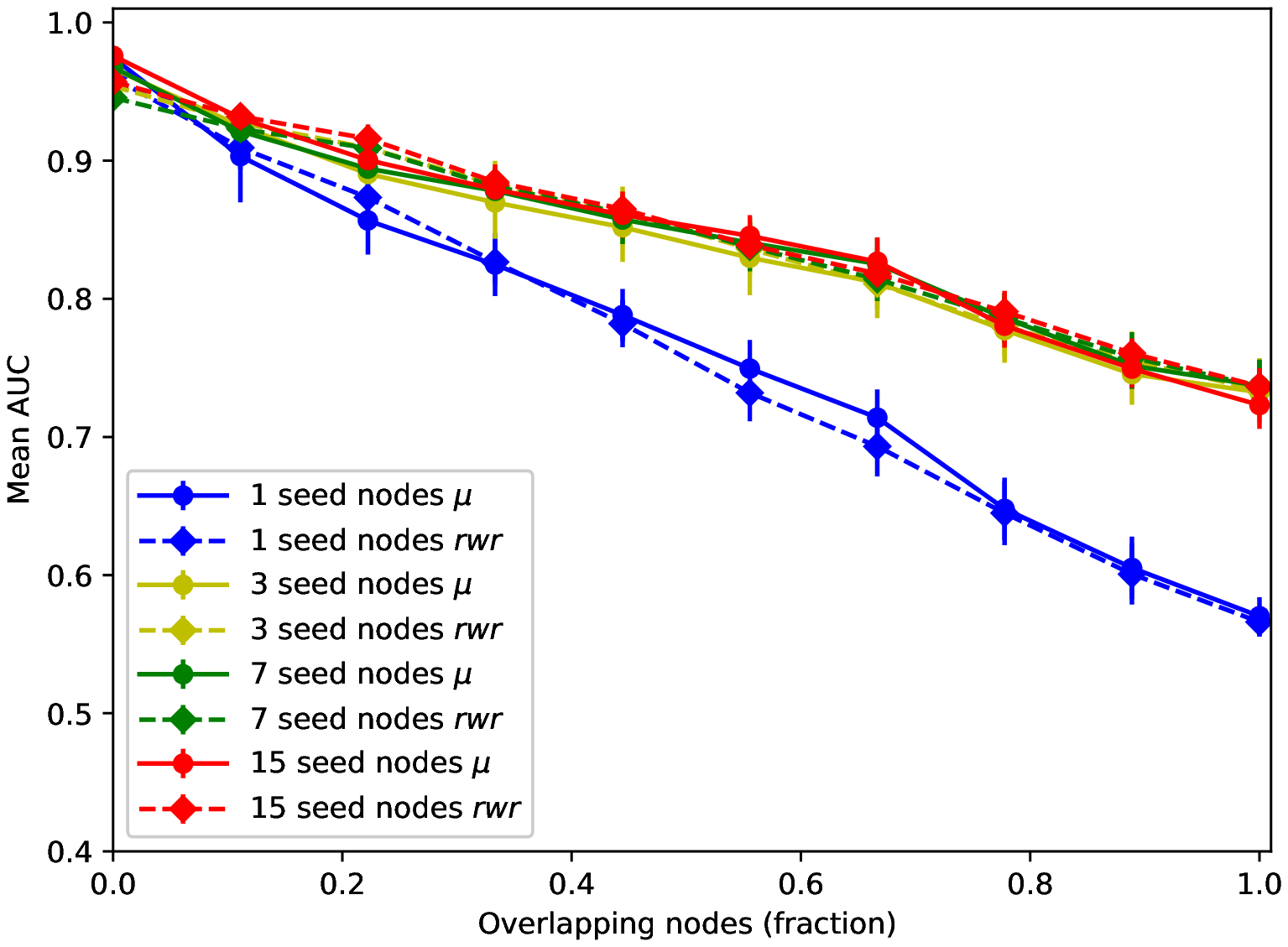}
        \caption{5000 nodes}
    \end{subfigure}
    \caption{Overlapping LFR community networks with varying seed nodes with 1000 and 5000 nodes.
     Data points represent mean AUC scores for all communities on 10 sampled networks at varying overlapping fraction of nodes.
     Error bars represent one standard deviation.}
     \label{fig:auc_overlap}
\end{figure*}

In Figure \ref{fig:auc_overlap} we show the results of network performance when tested against an increasing level of overlapping communities.
For these tests we fix the mixing coefficient at $0.3$.
Here, each vertex can belong to up to 4 communities.
In order to test performance we varied the fraction of nodes that are in more than one community.
The method still has an AUC score above 0.5 when all nodes are placed in multiple communities.
This indicates that the method is capable of uncovering latent overlapping memberships even when given a relatively small number of seed nodes.
As with the non-overlapping results, the scores are comparable with the rwr method with performance slightly improved in the case of 1000 node models, but comparable for 5000 node models.

\subsection{Real networks}
\label{sec:real_networks}

\begin{figure*}
     \centering
    \begin{subfigure}[b]{0.45\textwidth}
        \centering
        \includegraphics[width=\textwidth]{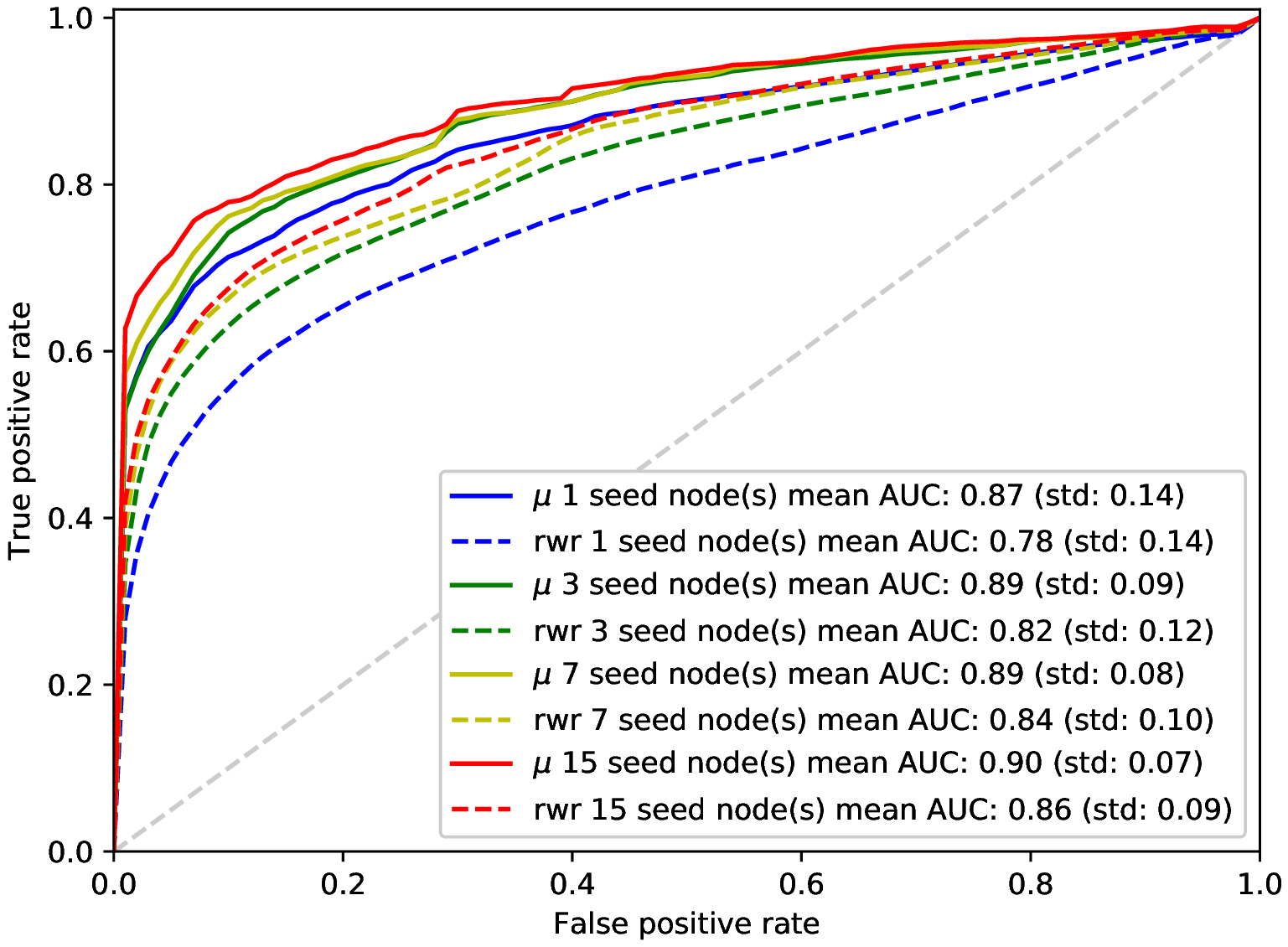}
        \caption{EU email departments}
    \end{subfigure}
    \begin{subfigure}[b]{0.45\textwidth}
        \centering
        \includegraphics[width=\textwidth]{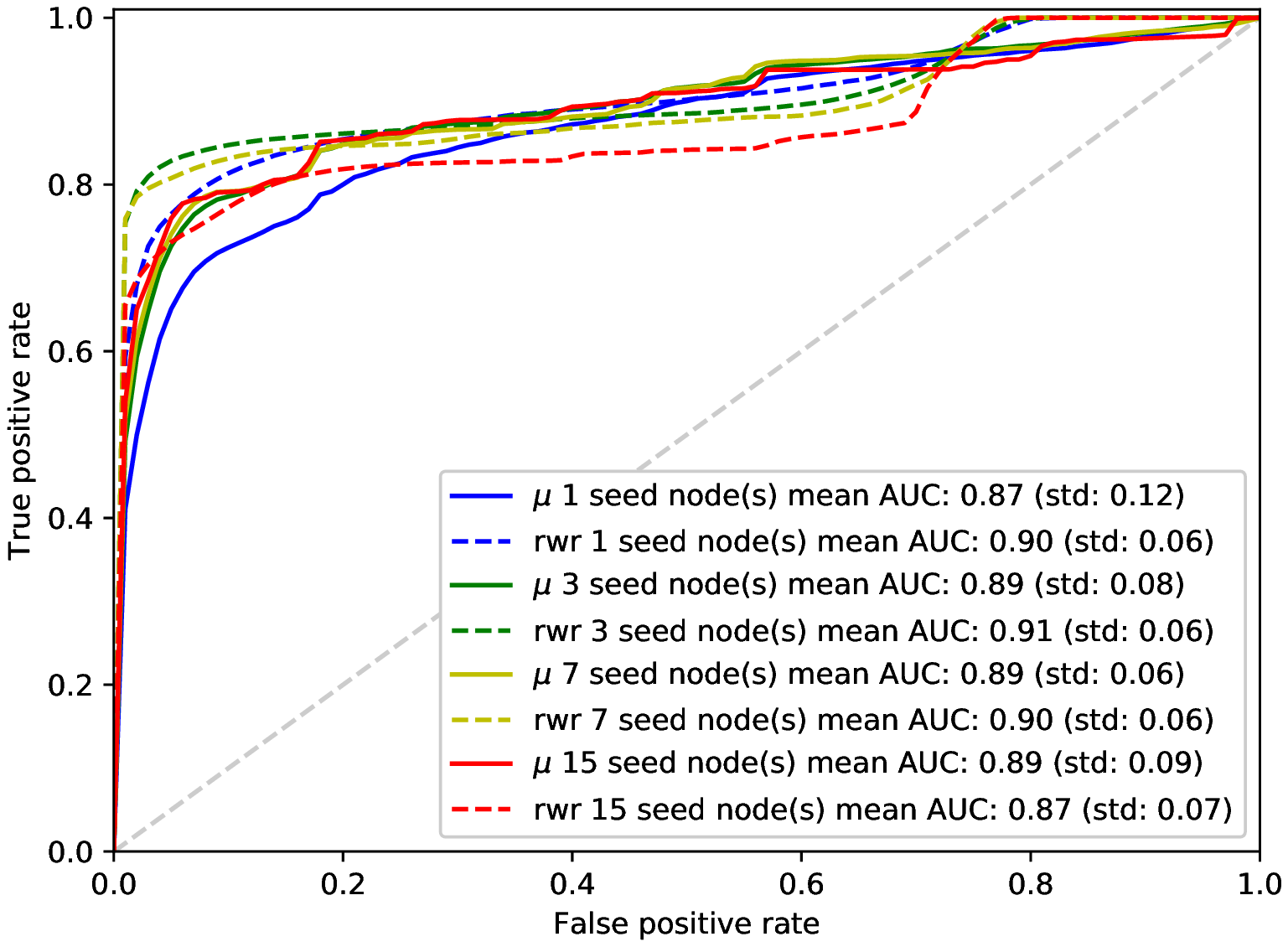}
        \caption{Yeast PPI protein complexes}
    \end{subfigure}
    
    \begin{subfigure}[b]{0.45\textwidth}
        \centering
        \includegraphics[width=\textwidth]{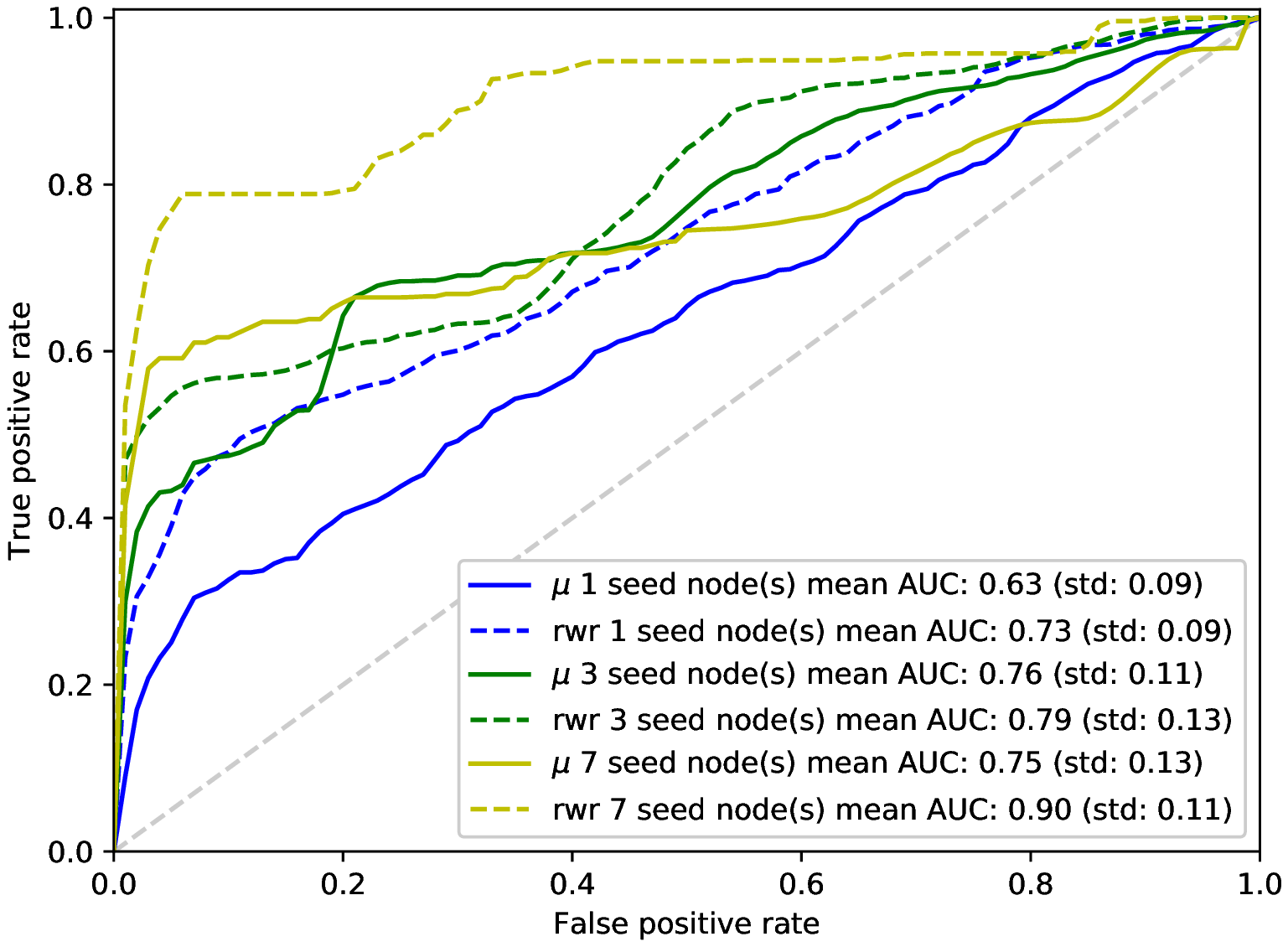}
        \caption{Arabidopsis PPI protein complexes}
    \end{subfigure}
    \begin{subfigure}[b]{0.45\textwidth}
        \centering
        \includegraphics[width=\textwidth]{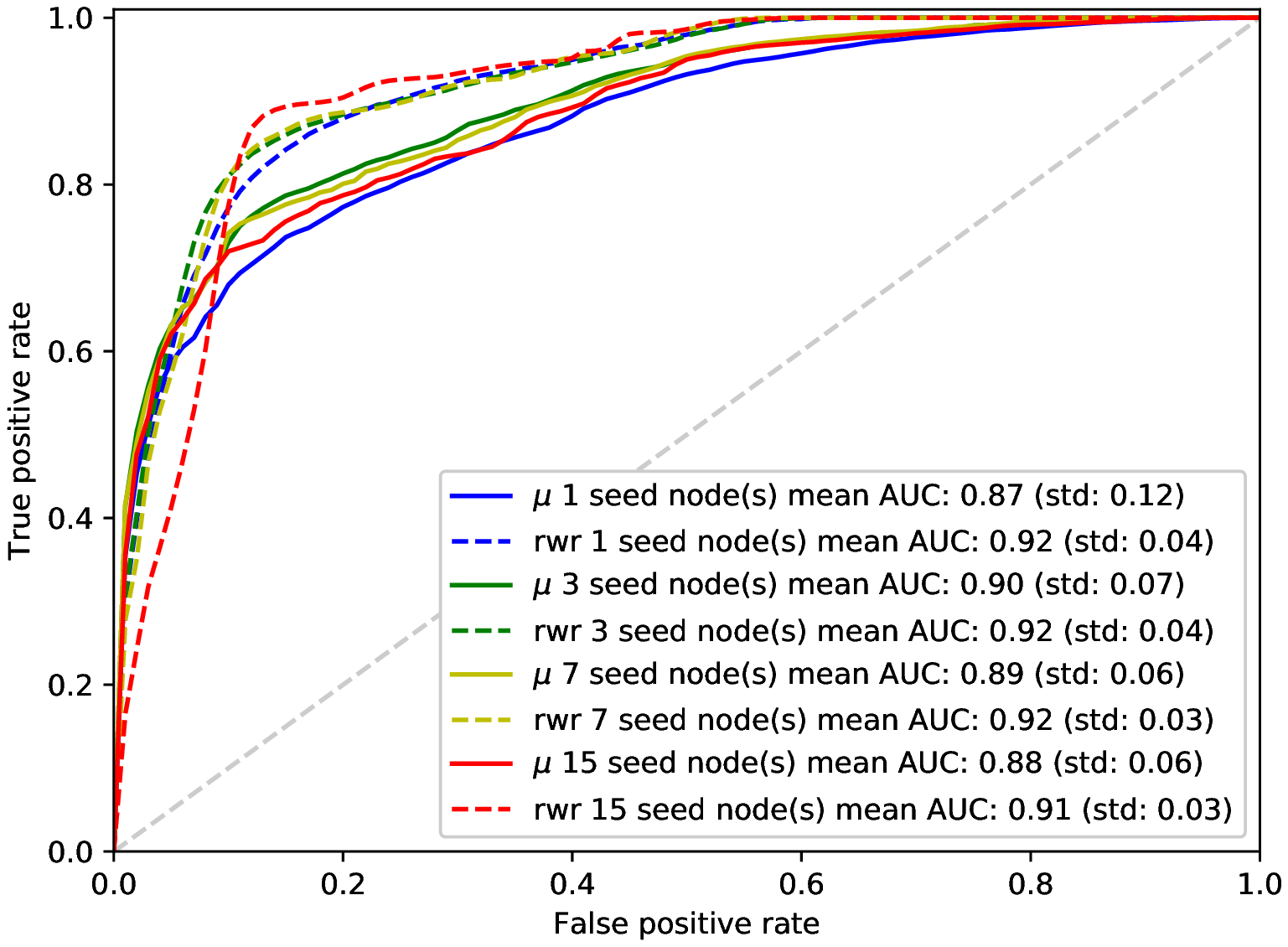}
        \caption{E. coli PPI protein complexes}
    \end{subfigure}
    \caption{Receiver operator characteristic (ROC) curves for the high quality labels for real networks described in Section \ref{sec:real_networks}. 
    Results represent the averages over all receiver operating characteristic scores from all samples of label sets with the cross validation procedure described in \ref{sec:cross_validation}.}
    \label{fig:quality_label_rocs}
\end{figure*}

In order to test the performance of the semi-supervised classification on real-world data we present our findings on example networks with metadata communities.
All datasets use the largest single connected component sub-graph.
The real networks used are:
\begin{itemize}
 \item \textbf{EU emails core dataset (EU emails) \cite{leskovec2007graph}} This anonymised dataset is taken from the SNAP database \cite{snapnets} and contains $986$ nodes and $16,687$ edges representing emails between individuals.
 The metadata community labels represent different departments within the organisation.
 In total there are 42 communities, 39 of which contain at least 3 nodes.
 
 \item \textbf{Yeast protein-protein interaction network (Yeast PPI)} \cite{yu2008high}
 This dataset is a collection of recorded binary interactions between proteins collected with high-throughput yeast-2-hybrid assays.
 The metadata used are known, experimentally validated protein complexes from \cite{pu2008up}.
 The network contains 6222 nodes in the largest connected component, with 22,386 edges.
 There are 409 experimentally validated protein complexes, 236 of which contain 3 or more nodes.
 The protein complexes are typically very small in terms of number of proteins, with  90\% of the complexes containing less than 10 proteins and only 2 complexes containing 50 or more proteins.

  \item \textbf{\textit{Escherichia coli} protein-protein interaction network (E.coli PPI) \cite{su2007bacteriome}}
  For the \textit{E. coli} dataset we used manually curated interacting proteins from \cite{su2007bacteriome}.
  The network contains 1913 nodes and 7252 edges, the protein complexes range in size between 3 and 65 nodes.
  85\% of the complexes contain 10 or less nodes.
 
 \item \textbf{\textit{Arabidopsis thaliana} protein-protein interaction network (Arabidopsis PPI)} \cite{ArabidopsisConsortium2011}.
  The network itself contains 4519 nodes and 11,096 edges.
  For the Arabidopsis dataset, the complex sources were more limited.
  Consequently, we obtained all gene ontology annotations under the GO term \textit{``Protein-containing complex''} from AmiGO \cite{carbon2009amigo} where experimentally collected physical interaction evidence was acquired.
  At the time of writing, this resulted in 7 complexes containing between 4 and 12 nodes.
  In addition, we included small protein complexes from the IntAct database \cite{orchard2013mintact}, resulting in a total of 165 unique complexes.
  As the labels for this dataset are small it was not possible to test algorithm performance with 15 seed nodes.
  
\end{itemize}

ROC curves generated with the cross validation procedure described in Section \ref{sec:cross_validation} are shown in Figure \ref{fig:quality_label_rocs}.
These results represent ROC curves and mean AUC statistics.

The results of our method are comparable to the random walk with restart approach, which performs better on the datasets tested with the exception of the EU email dataset for which our model produces higher AUC scores.

In the case of the \textit{Arabidopsis thaliana} network the protein complexes tested are significantly smaller than for other networks and so a comparison of seed sizes is not possible.
However, in the other networks studied, using 3 or more seed nodes appears to improve results, though for 15 seed nodes the results are not significantly better than with 7 seeds.
The lower quality labels within the Arabidopsis protein complex dataset likely explains the significant difference in results when compared with other protein interaction datasets.

\subsubsection{Tests on gene ontology labels}
\begin{figure*}[t]
     \centering
    \begin{subfigure}[b]{0.31\textwidth}
        \centering
        \includegraphics[width=\textwidth]{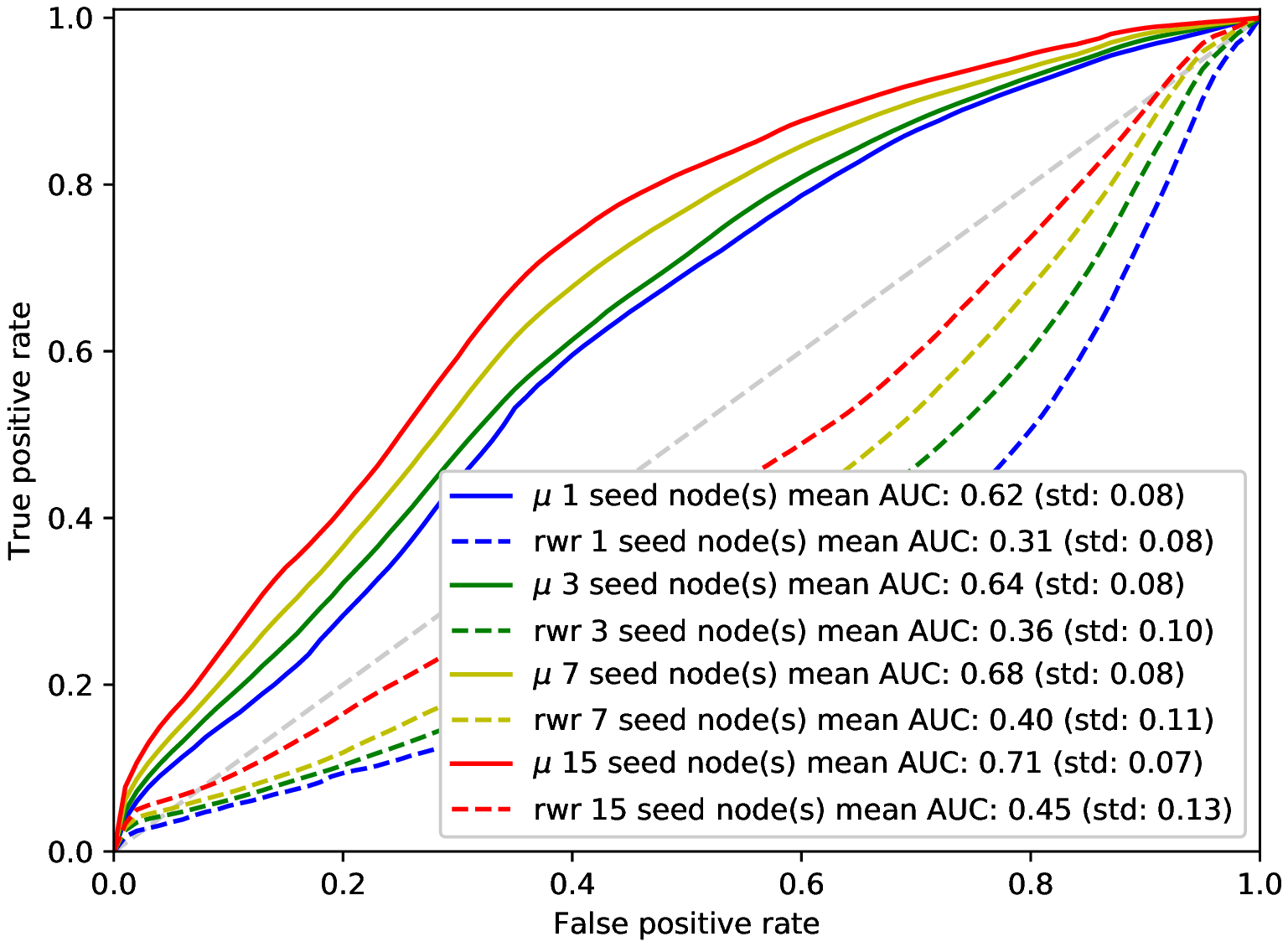}
        \caption{Yeast PPI GO terms}
    \end{subfigure}
    \begin{subfigure}[b]{0.31\textwidth}
        \centering
        \includegraphics[width=\textwidth]{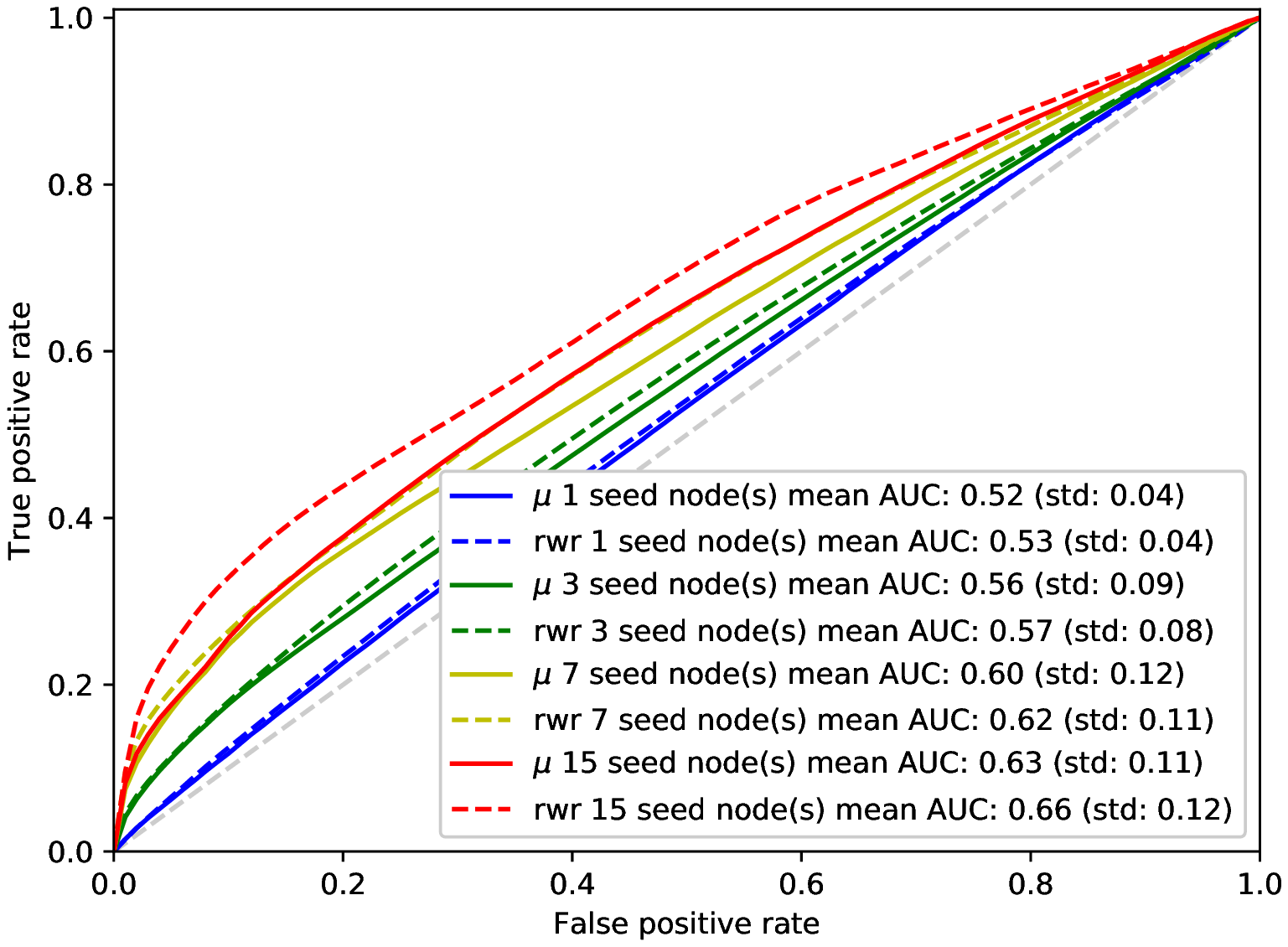}
        \caption{Arabidopsis PPI GO terms}
    \end{subfigure}
    \begin{subfigure}[b]{0.31\textwidth}
        \centering
        \includegraphics[width=\textwidth]{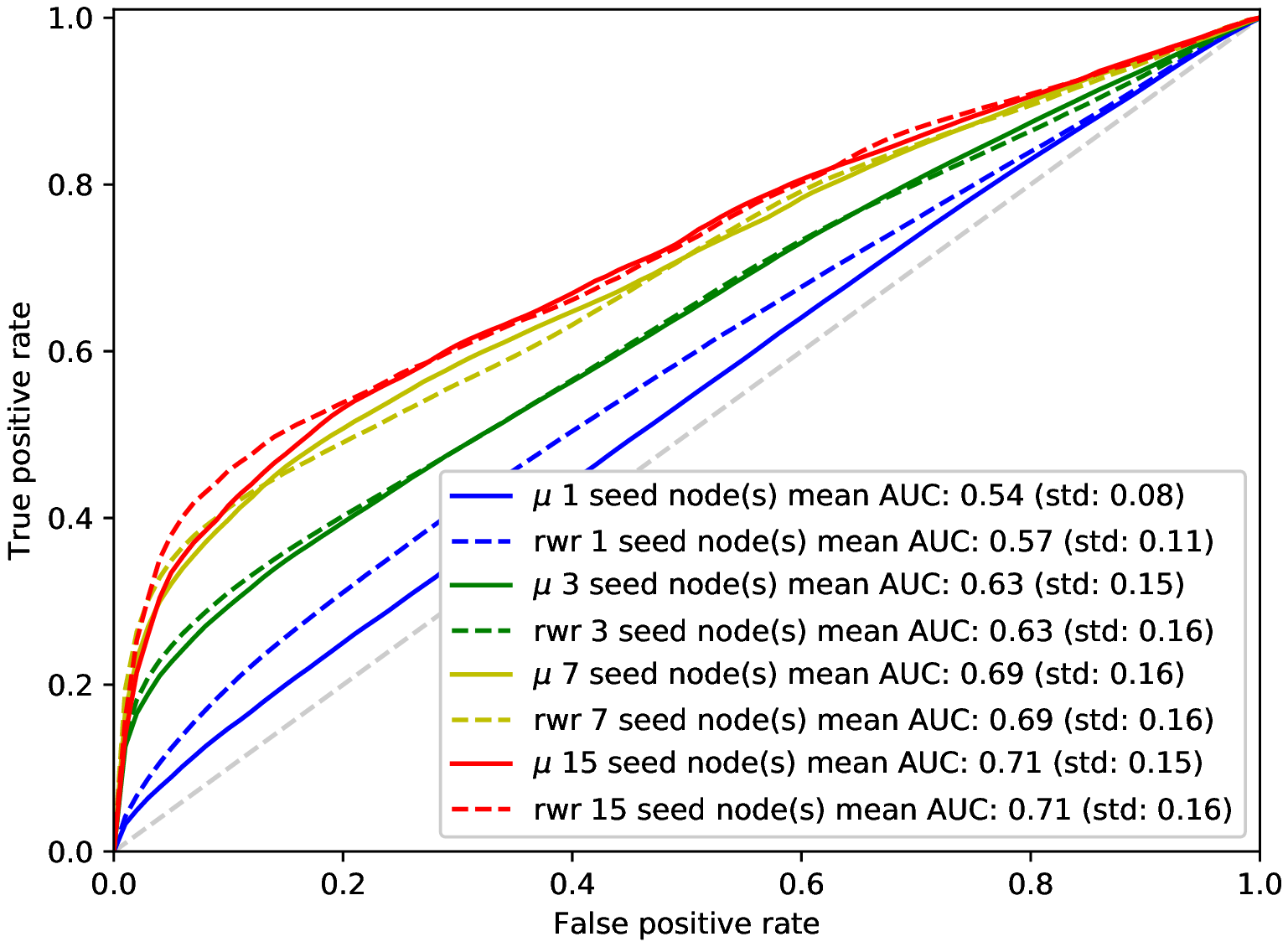}
        \caption{E. coli PPI GO terms}
    \end{subfigure}
    \caption{Receiver operator characteristic (ROC) curves for gene ontology labels for real networks described in Section \ref{sec:real_networks}. 
    Results represent the averages over all receiver operating characteristic scores from all samples of label sets with the cross validation procedure described in \ref{sec:cross_validation}.}
    \label{fig:GO_TERM_ROCS}
\end{figure*}


In a practical situation the higher quality label sets described above are not likely to be available.
Consequently, we wish to highlight that the method also serves as a useful network reduction tool where the quality of labels is not well defined.
In order to achieve this goal, for each of the biological data sets in this study, we acquired all Gene Ontology terms for each of the proteins from AmiGO \cite{carbon2008amigo}.
Gene Ontology is a controlled vocabulary of terms associated with biological functions in three broad categories of Cellular Components (CC),  Metabolic Functions (MF) and Biological Processes (BP), forming a hierarchy of terms and associated sub terms.

These labels are not all likely to be represented within the protein-protein interaction networks, though many biological processes are.
For the Yeast, \textit{Arabidopsis} and \textit{E. coli} PPIs we collected 85, 549 and 420 terms covering at least 3 nodes, respectively.
As expected, average ROC and AUC scores shown in Figure \ref{fig:GO_TERM_ROCS} are considerably lower than for the Protein complexes for both our method and the rwr approach. 
Notably, however, the performance of the rwr approach is significantly worse than one would expect to find at random for the Yeast GO terms.
The reason for this performance remains unknown, however, it is likely to to the fact that the gene ontology labels are poorly represented in the dataset.
However, this approach should be considered in contrast to conventional gene enrichment strategies on conventional network clusters \cite{chen2013enrichr}.

\section{Discussion}
The semi-supervised method for vertex classification presented in this work has been shown to produce good results on both synthetic benchmarks and real-world datasets.
Interestingly, this method is capable of correctly classifying communities with only a small number of seed query vertices.
These results show that this query method could be used as a powerful exploratory tool in network analysis.

The fact that the method is able to uncover small protein complexes seems to contradict the principle that modularity maximisation algorithms have a resolution limit \cite{fortunato2007resolution}.
Whilst this appears to be the case it is important to note that the resolution limit applies to a single partition of space.
Further work is needed to investigate why small communities are still detectable.
However, we speculate that it is likely due to the fact that the co-classification of vertices between different partitions remains fairly tolerant to changes.
In other words, the small cluster of nodes is always clustered in the same community, regardless of the partition.
We do note that, where communities are very small, any approach will be extremely sensitive to false positive and false negative results.
As such, this should be considered when using any method of this form as an exploratory tool.

The approach also appears to be tolerant to a small number of seed nodes.
This is interesting as in most sampled cases the relevant nodes are unlikely to be direct neighbours.
From the perspective of exploratory studies, this implies that a small number of query vertices can be used to find potentially related vertices.

\section{Related work}
\label{sec:related_work}
This work relates very strongly to the idea of local community detection, more specifically the idea of \textit{seed set expansion} \cite{gleich2012vertex}.
Here, a given seed set is created and random walks are analysed to find clearly related communities of vertices.
One of the most common approaches to finding related vertices in a network is the random walk with restart (RWR) \cite{can2005analysis, kohler2008walking} explained in section \ref{sec:rwr}.
This approach has been applied in fields as diverse as recommender systems and the detection of potential drug targets \cite{chen2012drug}.
In RWR the relatedness of any pair of vertices can be seen as the probability of a particle traversing the graph starting at a given vertex and ending at another.
Conceptually, RWR is very similar to the method presented in this paper given that the user has a given query.
The RWR probability as analogous though not equivalent to the value of $\mu_i(S)$.

The reader may also consider the similarity of this approach to that of label propagation \cite{raghavan2007near, gregory2010finding}, which seeks to find communities based on a vote, where at different time steps a node updates it's label to to be the most common amongst its neighbours.
A common problem with label propagation schemes is that they often fail to converge, resulting in many competing clusterings of a network.
We liken this to the problem encountered by Good \textit{et al.} \cite{good2010performance} in modularity maximisation.
A core contribution of this paper is that, in such situations, there is no single, context independent labelling scheme that can be seen as the ``true'' community structure, overlapping or not.
Indeed, the approach applied here is extremely general and could be adapted to the label propagation approach (or any other community detection algorithm) should sufficient semi-supervised group memberships be known \textit{a priori}.

Many existing method are based on the idea of a locally dense subgraph containing \textit{all query nodes} \cite{benson2016higher}.
In contrast, the query approach presented here does not require the queries to be a self contained sub-graph.
Indeed, queries can contain spurious nodes that are topologically distant from one another - the result is that the $\mu_i(S)$ score for the query set will likely be very low.
Further investigation into how to evaluate the quality of high average $\mu_i(S)$ for $i \in S$ is left to future work.

In the field of community detection, a number of very recent articles have focused on using metadata to improve the results of community detection approaches.
These algorithms, however, are distinct from the approach taken here as the metadata is not used in the module discovery process.
Furthermore, the results in this work attempt to explicitly label unlabelled data and only require a relatively small number of labels to operate in such a fashion.
In contrast,  the recent approach by Newman and Clauset \cite{newman2016structure}, for example, uses examples in which practically the entire network contains labels which is less useful from the perspective of label discovery.

\section{Conclusions}
This paper has presented a novel approach to semi-supervised community detection utilising a consensus of high scoring partitions computed with the popular modularity maximisation approach.
Previously the glassy search space of this optimisation algorithm has been seen as a major limitation.
However, in this work we consider each locally optimal partition to be information regarding the true multi-class labels that are likely present in real networks.
The approach presented here differs from other ensemble approaches in that the objective is to provide a probabilistic framework for label classification.
Performance was shown to be strong on both synthetically generated networks and real-world ground truth communities with relatively small sets of labels.
In the case of synthetic networks communities are correctly detected up to the detectability threshold.
For real world networks with small label sets, average AUC scores were comparable to the random walk with restart method for the high quality datasets tested in this study.

However, this approach requires the community landscape to contain many local maxima, a property likely shared by many real-world, heterogeneous networks.
Similarly, the method presented here requires both some labelled data and the labels to be relevant in the context of the underlying network.

This work presents a number of interesting potential future avenues for research, such as observing how query sets change in time dynamic or multi-scale networks.
Furthermore, as the algorithm is trivial to run in a distributed manner, this approach could be applied to larger graphs than those studied in this paper.
Further research should also be conducted into how this approach could be applied to other partition quality functions, such as the infomap algorithm \cite{Rosvall:2008fi}.
The method was implemented in python and all software is available at: \\
\url{https://github.com/SBRCNottingham/cluster_query_tool}.

\section*{Acknowledgements}
We would like to thank Nicole Pearcy and Jenna Reps for assistance preparing this manuscript.
We are grateful for access to the University of Nottingham High Performance Computing Facility.
This work was supported by the UK Biotechnology and Biological Sciences Research Council (BBSRC) grant BB/L013940/1.

\bibliographystyle{ACM-Reference-Format} 
\bibliography{references}

\end{document}